\documentclass[10pt,conference]{IEEEtran}
 
\UseRawInputEncoding
\usepackage{cite}
\usepackage{amsmath}
\usepackage{amsfonts}
\usepackage{graphicx}
\usepackage{float}
\usepackage[table]{xcolor}
\usepackage{subfigure}
\usepackage{epsfig}
\usepackage{amsthm}
\usepackage{mathrsfs}
\usepackage{amsfonts}
\usepackage{amssymb}
\usepackage{mathrsfs}
\usepackage{euscript}
\usepackage{multirow}
\usepackage{url}
\usepackage{algorithm}
\usepackage{algorithmic}
\usepackage{bm}
\usepackage{cuted}
\usepackage{tcolorbox}
\usepackage[font=footnotesize]{subcaption}
\usepackage[font=footnotesize]{caption}
\usepackage{flushend}

\usepackage{hyperref}
\usepackage{xcolor, soul}
\sethlcolor{green}

\usepackage{tikz}

\newcommand\submittedtext{
  \footnotesize \textcopyright{} 2025 IEEE. Personal use of this material is permitted. Permission from IEEE must be obtained for all other uses, in any current or
future media, including reprinting/republishing this material for advertising or promotional purposes, creating new collective works,
for resale or redistribution to servers or lists, or reuse of any copyrighted component of this work in other works. This is the accepted
version, not the final IEEE version. For IEEE version refer to DOI: \href{https://doi.org/10.1109/WCNC61545.2025.10978826}{10.1109/WCNC61545.2025.10978826}.}

\newcommand\submittednotice{
\begin{tikzpicture}[remember picture,overlay]
\node[anchor=north,yshift=-8pt] at (current page.north) {\fbox{\parbox{\dimexpr0.99\textwidth-\fboxsep-\fboxrule\relax}{\submittedtext}}};
\end{tikzpicture}
}

\renewcommand\fbox{\fcolorbox{red}{white}}
\theoremstyle{remark}

\newtheorem*{remark*}{Remark}

\DeclareMathOperator{\E}{\mathbb{E}}                     % Expectation

\begin{document}
\bstctlcite{IEEEexample:BSTcontrol}
\title{Rician Channel Modelling for Super Wideband MIMO Communications
\vspace{-3mm}
}

\author{Sachitha C. Bandara$^*$, Peter J. Smith$^\dagger$, Erfan Khordad$^*$, Robin Evans$^*$,  Rajitha Senanayake$^*$\\
$^*$ Department of Electrical and Electronic Engineering, University of Melbourne, Melbourne, Australia\\
$^\dagger$ School of Mathematics and Statistics, Victoria University of Wellington, Wellington, New Zealand
\vspace{-2mm}
}

\maketitle
\submittednotice
\begin{abstract}

Recent developments in Multiple-Input-Multiple-Output (MIMO) technology include packing a large number of antenna elements in a compact array to access the bandwidth benefits provided by higher mutual coupling (MC). The resulting super-wideband (SW) systems require a circuit-theoretic framework to handle the MC and channel models which span extremely large bands. 
Hence, in this paper, we make two key contributions. First, we develop a physically-consistent Rician channel model for use with SW systems. Secondly, we express the circuit-theoretic models in terms of a standard MIMO model, so that insights into the effects of antenna layouts, MC, and bandwidth can be made using standard communication theory. For example, we show the bandwidth widening resulting from the new channel model. In addition, we show that MC distorts line-of-sight paths which has beamforming implications. We also highlight the interaction between spatial correlation and MC and show that tight coupling reduces spatial correlations at low frequencies.

\end{abstract}
 
 % Proper MC modelling can transform MC from a traditionally adverse effect to a key enabler of massive bandwidths.To integrate MC into communication theory, multiport network theory, which models end-to-end networks as multiport circuits is employed.
\begin{IEEEkeywords}
 Mutual Coupling, Super Wideband, Channel Modelling, Rician Channels, Spatial Correlations, Frequency Correlations
\end{IEEEkeywords}
 
 \vspace{-3mm}
\section{Introduction}\label{Sec:Introduction}

The advantages of conventional Multiple-Input-Multiple-Output (MIMO) in terms of enhanced spatial diversity and spectral efficiency are well-known \cite{chataut2020massive}. More recently, %his trend has recently evolved into using many antenna elements on the antenna array, resulting in 
interest has turned to the use of large arrays in compact spaces resulting in closely packed antenna elements. In addition to increasing spatial diversity and spectral efficiency, the use of closely packed antennas %this trend positively impacts the operational bandwidth of the antenna array, 
leads to operational bandwidth widening
 and \textit{super-wideband} (SW) systems \cite{super_wideband}.

%The operational bandwidth widening of closely spaced antenna elements 
This bandwidth widening effect has been demonstrated in \cite{super_wideband} through modelling the mutual coupling (MC) between antenna elements in an array. Although traditional array designs typically avoid the adverse effects of high MC by spatially isolating antenna elements \cite{mutual_coupling_2018, wallace2004}, designs of colinear connected arrays have been shown to achieve SW performance \cite{munk2006connected, connected_arrays}. However, previous works are heavily based on circuit theory and the use of simple channel models when analyzing communications performance. Also, the channel models used do not maintain consistency in the parameters and channel values across the massive bandwidth.

Hence, in this paper,  we make two
key contributions. First, we develop a physically-consistent, correlated Rician channel model for use across the whole bandwidth.  Secondly, we express
the circuit-theoretic models in terms of a standard MIMO model,
so that insights into the effects of antenna layouts, MC and
bandwidth can be made using standard communication theory.

In developing the correlated Rician channel model, we integrate multiport network theory \cite{Ivrlac_main} into communication theory to properly account for MC effects and provide a physically-consistent channel model. We include spatial channel correlations and frequency correlations in our channel modelling, as opposed to \cite{super_wideband, matlab_mc_model} that only considered spatial correlations. We present a method to compute the Rician K-factor across the frequency domain based on recent measurement campaigns. Numerical results based on the developed channel model show the SW gains of tightly coupled arrays in the presence of spatial and frequency correlations. We also show two key changes to the global channel structure caused by tightly coupled arrays: a) the steering vector structure of line-of-sight (LoS) rays is distorted and has beamforming implications; and b) the spatial correlation at low frequencies is reduced. All of these observations are illustrated using extensive numerical examples. 

% \textcolor{blue}{(Erfan: can we put this [8] into one of the earlier sentences? cuz this is the first time we mention it)}
\noindent
\textit{Notation}: $\Re\{.\}$ represents the real part of a complex number or matrix. For any matrix $\mathbf{A}$ or vector $\bm{a}$, $\mathbf{A}^T$ and $\bm{a}^T$ denote the transpose, while $\mathbf{A}^H$ and $\bm{a}^H$ indicate the conjugate transpose. $\text{diag}(\mathbf{A})$ creates a diagonal matrix by setting all off-diagonal elements to zero. $\mathbf{I}_M$ is the identity matrix of size $M \times M$, and $\mathbb{E}[.]$ denotes the statistical expectation.

\section{System Model with Circuit Theory framework}\label{Sec:systemModel}
% \subsection{Circuit Theory for MIMO Communications}\label{ssec:cct_theoretic_framework}
Let us first consider a system model where we have a transmitter with $N_t$ antennas transmitting to a receiver with $N_r$ antennas in a multipath fading environment. In a traditional setting, the resultant MIMO channel can be modelled using classical communication theory. However, in this work, we aim to explore how MC impacts the system performance. To facilitate this analysis, we adopt a circuit-theoretic approach, as introduced in \cite{Ivrlac_main}, which provides a more detailed and physically-consistent representation of MC effects. Using input-output voltage relationships, we write the resultant MIMO system equation as given below,
\begin{align}
    \bm{v}_{L}(f) &= \mathbf{H}(f)\bm{v}_{G}(f) + \bm{n}(f),
    \label{eqn:system_eqn}
\end{align}
where $\bm{v}_{L}(f) \in \mathbb{C}^{N_r\times 1}$ and $\bm{v}_{G}(f)\in \mathbb{C}^{N_t\times 1}$ are the frequency domain representations of load (output) and source or generator (input) voltage vectors, respectively. The term $\mathbf{H}(f) \in \mathbb{C}^{N_r\times N_t}$ denotes the equivalent MIMO channel matrix, and $\bm{n}(f) \in \mathbb{C}^{N_r\times 1}$ represents the noise voltage vector at the receiver. We use frequency-domain representations throughout this paper as they provide insights into how the system behaves at different frequencies across the wide frequency bandwidth. In this setup, the MIMO channel matrix can be modelled using the self and mutual impedances of the antenna arrays at both ends and the parameters of other linearly connected devices such as low noise amplifiers (LNA) \cite{super_wideband}. For an end-to-end MIMO system with an LNA structure, where each LNA has a gain $\beta$ and internal resistance $R_{in}$, we can write the equation for $\mathbf{H}(f)$ as \cite{super_wideband}, 
\begin{subequations}
    \begin{align}
    \mathbf{H}(f) &= \beta R_{in}\mathbf{P}(f)\mathbf{Z}_{RT}(f)\mathbf{Q}(f),
    \end{align}
    \noindent where,
    \begin{align}
    \mathbf{P}(f) &= \left(\mathbf{Z}_{R}(f) + R_{in}\mathbf{I}_{{N}_{r}}\right)^{-1}, \\
    \mathbf{Q}(f) &= \left(\mathbf{Z}_{T}(f) + R\mathbf{I}_{{N}_{t}}\right)^{-1}.
    \end{align}
    \label{eqn:matrix_relationships_mimo}
\end{subequations}
\vspace{-1em}
\par\noindent In (\ref{eqn:matrix_relationships_mimo}), $R$ is the resistance of source voltage generators, $\mathbf{P}(f) \in \mathbb{C}^{N_r \times N_r}$ and $\mathbf{Q}(f) \in \mathbb{C}^{N_t \times N_t}$ are the receiver and transmitter coupling matrices, respectively. Note that the above equation is written for far-field (FF) communications. $\mathbf{Z}_{T}(f) \in \mathbb{C}^{N_{t}\times N_{t}}$ and $\mathbf{Z}_{R}(f) \in \mathbb{C}^{N_{r}\times N_{r}}$ are the transmitter and receiver array impedance matrices, respectively. Both matrices are square symmetrical matrices in which the diagonal elements correspond to self impedances, while the off-diagonal elements denote mutual impedances between elements. $\mathbf{Z}_{RT}(f) \in \mathbb{C}^{N_r \times N_t}$ is the transmit-receive trans-impedance matrix which captures the pairwise impedance between the elements of transmitter and receiver arrays\footnote{Here, we ignored $\mathbf{Z}_{TR}(f) \in \mathbb{C}^{N_t \times N_r}$, the receive-transmit trans-impedance matrix as in \cite{Ivrlac_main, super_wideband} using the \textit{unilateral approximation} for far-field (FF) communications.}. This trans-impedance matrix models the electromagnetic wave propagation characteristics in the circuit theoretic-approach for FF communications. It can be written as,
\begin{align}
    &\mathbf{Z}_{RT}(f) \nonumber \\&= \text{diag}(\Re{\{\mathbf{Z}_R (f)\}})^{\frac{1}{2}} \mathbf{H}_{\text{MIMO}}(f) \text{diag}(\Re{\{\mathbf{Z}_T (f)\}})^{\frac{1}{2}} e^{j\phi},
    \label{eqn:cct_theoretic_channel}
    \vspace{-2em}
\end{align}
where \(\phi\) represents a frequency-dependent circuit-theoretic parameter specific to the equivalent RLC circuit of an antenna element. We will discuss the specific relationship used for $\phi$ in our study in Section \ref{Sec:Simulations}. Further, $\mathbf{H}_{\text{MIMO}}(f)$ is the conventional MIMO channel matrix that we model using communication theory. For SW systems, it requires careful design to ensure consistency of parameters and channel behavior across the massive bandwidth. Calculation of the diagonal elements of $\mathbf{Z}_{T}$ and $\mathbf{Z}_{R}$ are straightforward and can be done using the equivalent electrical circuits of the corresponding antennas used in the arrays. For the off-diagonal elements we adopt a closed-form formula for mutual impedance between two small antennas under the \textit{Chu limit}  which was derived in \cite{shyianov2022, super_wideband}.

Now let us focus on the noise. Compared to traditional channel modelling, where we consider receiver noise to be white and Gaussian, here we adopt a circuit framework that considers physically-consistent noise sources \cite{Ivrlac_main}. As such, the received noise vector in (\ref{eqn:system_eqn})  for FF communications can be written as \cite{super_wideband},
\begin{align}
    \bm{n}(f) = \bm{v}_{N}(f) + \beta R_{in}\mathbf{P}(f)\widetilde{\bm{v}}_{N,R}(f),
    \label{eqn:noise_equation}
\end{align}
where $\bm{v}_{N}(f) \in \mathbb{C}^{N_r \times 1}$ is the noise voltage vector produced by LNAs and $\widetilde{\bm{v}}_{N,R}(f) \in \mathbb{C}^{N_r \times 1}$ is the noise voltage vector produced by the receiver antenna array due to thermal agitation. In (\ref{eqn:noise_equation}), noise voltages at the transmitter array are ignored as they are dominated by a much larger transmit power. To analyze the system performance, we need to examine second-order statistics of these noise voltages. When MC is taken into account, these complex noise voltages at both the transmitter and the receiver are correlated. However, for FF communications, when MC effects can be ignored between the transmitter and the receiver arrays, the transmit-receive noise voltages are uncorrelated. Therefore, the noise covariance matrix at the receiver array originating from receiver antennas can be expressed as,
\begin{equation}
    \E\left[{\Tilde{\bm{v}}_{N,R}(f)}{\Tilde{\bm{v}}_{N,R}^{H}(f)}\right] = 4k_{b}T\Delta f \Re{\{\mathbf{Z}_R(f)\}},
\end{equation}
where $k_b$ is the Boltzmann constant, $T$ is the absolute room temperature, and $\Delta f$ is the channel bandwidth. In a multi-carrier communication system, $\Delta f$ can be treated as the bandwidth of an equally spaced frequency sub-channel. For the receiver LNA structure with each having an internal resistance of $R_{in}$ and a noise figure $N_{f}$, the noise covariance matrix can be expressed as\footnote{Here, the noise correlation model does not incorporate noise coupling of externally connected devices such as LNAs, however, these are known to couple through the antenna array. For a rigorous review, refer to \cite{warnick2009}.},
\begin{equation}
    \E\left[{\bm{v}}_{N}(f){\bm{v}}_{N}^{H}(f)\right] = 4k_{b}T\Delta f R_{in}(N_{f} - 1)\mathbf{I}_{N_r},
\end{equation}
where $R_{in}(N_{f} - 1)$ is the equivalent thermal noise resistance of the LNA. Thus, for FF communications, the final covariance matrix of the received noise can be written as,
\begin{align}
    \mathbf{R}_{n}(f) &= \E\left[\bm{n}(f)\bm{n}^H(f)\right] = 4k_{b}T\Delta fR_{in} \left[(N_{f}-1)\mathbf{I}_{N_r} \right. \nonumber \\
    &\quad \left. + \beta^2R_{in}\mathbf{P}(f)\Re\{\mathbf{Z}_R(f)\}\mathbf{P}^{H}(f)\right].
    \label{noise_correlation}
\end{align}

\noindent For the communication scenario described in Section \ref{Sec:systemModel}, we will now discuss how the channel matrix, $\mathbf{H}_{\text{MIMO}}(f)$ is modelled.
\vspace{-2mm}
\section{Super Wideband Channel Model}
In this work, we consider a correlated Rician fading channel model and assume Uniform Linear Arrays (ULAs) in co-linear configuration at both the transmitter and the receiver. The $N_r \times N_t$ channel matrix can be expressed as,
\begin{align}
    \mathbf{H}_{\text{MIMO}}(f) &= \sqrt{\beta(f)} \left[ \sqrt{\frac{K(f)}{K(f)+1}}\bm{a}_R(\theta_R,f)\bm{a}_T^{H}(\theta_T,f) \right. \nonumber \\
    &\quad \left. + \sqrt{\frac{1}{K(f)+1}}\mathbf{R}_{R}^{1/2}(f)\mathbf{U}(f)\mathbf{R}_{T}^{1/2}(f) \right],
    \label{eqn:Rician_channel_model}
\end{align}
where $K(f)$ denotes the frequency dependent Rician K-factor and $\bm{a}_R(\theta_R,f)$, $\bm{a}_{T}(\theta_T,f)$ represent the receiver and transmitter steering vectors where $\theta_R$ and $\theta_T$ are the angles of arrival and departure, respectively. Further, $\mathbf{R}_{R}(f)$ and $\mathbf{R}_{T}(f)$ represent the receiver and transmitter spatial correlation matrices assuming the well-known Kronecker structure. The entries of $\mathbf{U}(f)$ are independent and identically distributed (i.i.d.) complex values drawn from a standard complex normal distribution. The term, $\beta(f)$ is the channel gain given by,
\begin{align}
    \beta(f) &= \beta_{LoS}(f)\left(1+1/K(f)\right),
    \vspace{-2mm}
\end{align}
where $\beta_{LoS}(f) = G_{T}G_{R}\left(c/(2\pi fd^{\gamma/2})\right)^2,$ with $G_T$ and $G_R$ representing the transmitter and receiver antenna gains, respectively. Further, $d$ is the distance between the transmit and receive arrays and $\gamma$ is the path loss exponent. 

The received noise voltage is correlated due to coupling, but uncorrelated noise is advantageous because it simplifies signal processing. For the remainder of this paper, we assume an uncorrelated noise model by whitening the output. To achieve this, we multiply (\ref{eqn:system_eqn}) by $\mathbf{R}_{n}^{-\frac{1}{2}}$, which effectively decorrelates the noise. This transformation not only simplifies the analysis but also allows us to isolate the effects of coupling solely within the channel matrix. The resultant whitened output voltage vector, denoted by $\Tilde{\bm{v}}_{L}(f)$, can be expressed as,
\begin{align}
    \Tilde{\bm{v}}_{L}(f) &= \widetilde{\mathbf{H}}(f)\bm{v}_G(f) + \Tilde{\bm{n}}(f),
    \label{eqn:whitened_mimo_sys}
\end{align} 
where the noise vector is $\Tilde{\mathbf{n}}(f) \sim \mathcal{C}\mathcal{N}(0,\mathbf{I}_{N_{r}})$. Assuming the antenna elements at one end have the same RLC parameters, their impedances are equal. Thus, let $\text{diag}(\Re\{\mathbf{Z}_T (f)\}) = \Re\{Z_1(f)\} \mathbf{I}_{N_t} \text{ and } \text{diag}(\Re\{\mathbf{Z}_R (f)\}) = \Re\{Z_2(f)\} \mathbf{I}_{N_r}$. Based on this, we can rewrite $\Tilde{\bm{v}}_{L}(f)$ as,
\begin{align}   
    \Tilde{\bm{v}}_{L}(f)  &= \beta R_{in}\sqrt{\Re{\{Z_{1}(f)\}}\Re{\{Z_{2}(f)\}}}e^{j\phi}\nonumber\\
    & \times \mathbf{R}_{n}^{-\frac{1}{2}}(f)\mathbf{P}(f)\mathbf{H}_{\text{MIMO}}(f)\mathbf{Q}(f)\bm{v}_{G}(f) + \Tilde{\mathbf{n}}(f).
\end{align} 
Let us closely inspect the LoS and the scattered channel components of the voltage vector. Defining $\alpha(f) = \beta R_{in}\sqrt{\Re{\{Z_{1}(f)\}}\Re{\{Z_{2}(f)\}}}e^{j\phi}$ and using (\ref{eqn:Rician_channel_model}), the equivalent LoS channel, $\mathbf{H}_{eq}^{LoS}(f)$ can be expressed as,
\begin{align}
    \mathbf{H}_{eq}^{LoS}(f)&= \alpha(f)\sqrt{\beta_{LoS}(f)}\bm{w}_{R}(\theta_R,f)\bm{w}_{T}^{H}(\theta_T,f),
    \label{eqn:h_los_equiv}
\end{align}
where,
\vspace{-2mm}
\begin{subequations}
    \begin{align}
    \bm{w}_{R}(\theta_R,f)&= \mathbf{R}_{n}^{-\frac{1}{2}}(f)\mathbf{P}(f)\bm{a}_{R}(\theta_R, f),\\
    \bm{w}_{T}(\theta_T,f)&= \mathbf{Q}^{H}(f)\bm{a}_{T}(\theta_T, f).
    \end{align}
    \label{eqn:equivalent_steering_vectors}
\end{subequations}
\noindent 
Note that the simple LoS channel structure involving only the steering vectors $\bm{a}_R(.)$, $\bm{a}_T(.)$, is scaled and distorted by coupling via $\mathbf{P}(f)$ and $\mathbf{Q}(f)$, and whitening via $\mathbf{R}_n(f)$. Similar to (\ref{eqn:h_los_equiv}), the equivalent scattered channel, $\mathbf{H}_{eq}^{Sc}(f)$ can be written as,
\vspace{-2mm}
\begin{align}
    \mathbf{H}_{eq}^{Sc}(f)&= \alpha(f)\sqrt{\frac{\beta(f)}{K(f)+1}}\mathbf{C}_{R}^\frac{1}{2}(f)\mathbf{U}(f)\mathbf{C}_{T}^\frac{1}{2}(f),
    \label{eqn:h_sc_equiv}
\end{align}
\vspace{-1mm}where,
\vspace{-1mm}
\begin{subequations}
    \begin{align}
    \mathbf{C}_{R}(f)&= \mathbf{R}_{n}^{-\frac{1}{2}}(f)\mathbf{P}(f)\mathbf{R}_{R}(f)\mathbf{P}^{H}(f)\mathbf{R}_{n}^{-\frac{1}{2}}(f),\\
    \mathbf{C}_{T}(f)&= \mathbf{Q}^{H}(f)\mathbf{R}_{T}(f)\mathbf{Q}(f),
    \end{align}
    \label{eqn:equivalent_spatial_correlations}
\end{subequations}
\vspace{-0.4cm}
\par\noindent can be thought of as equivalent receiver and transmitter spatial correlation matrices, respectively. From (\ref{eqn:h_sc_equiv}) and (\ref{eqn:equivalent_spatial_correlations}), it can be seen that the scattered components have exactly the same Kronecker structure as the scattered channel, but coupling has made substantial changes to the gain and the correlation structure. The final circuit theoretic and correlated Rician channel can now be written as,
\begin{align}
    \widetilde{\mathbf{H}}(f) &= \alpha(f)\sqrt{\beta(f)}\left[ \sqrt{\frac{K(f)}{K(f)+1}}\bm{w}_R(\theta_R,f)\bm{w}_T^{H}(\theta_T,f) \right. \nonumber \\
    &\quad \left. + \sqrt{\frac{1}{K(f)+1}}\mathbf{C}_{R}^{1/2}(f)\mathbf{U}(f)\mathbf{C}_{T}^{1/2}(f) \right].
    \label{eqn:sw_mimo_channel}
    \vspace{-2mm}
\end{align}
The channel model in (\ref{eqn:sw_mimo_channel}) converts the circuit theoretic models into an equivalent standard system model. Hence, we can gain insights into the effects of correlated noise and MC, simply by looking at the equivalent channels. 

Let us now consider the channel matrix in (\ref{eqn:sw_mimo_channel}) for SW systems. Any modelling of the SW MIMO system requires a channel model capable of spanning a wide frequency range maintaining consistency in the parameters and channel values across the band. Thus, in the following, we discuss how we model the spatial correlations, frequency correlations, and Rician K-factor in detail for SW MIMO systems.

\subsubsection{Spatial correlations}\label{ssec:spatial_corr}

To model the spatial correlations in the Kronecker model, $\mathbf{R}_R$ or $\mathbf{R}_T$, we use the local scattering spatial correlation model \cite{emil2017}, which assumes multipath components originating from a scattering cluster around the transmitter. The $ij^{th}$ entry of a generic spatial correlation matrix $\mathbf{R}_s$ can be represented as,
\vspace{-1mm}
\begin{align}
    [\mathbf{R}_s]_{ij} &= \int e^{j 2\pi \delta \frac{f}{c}(j-i)\sin({\Bar{\varphi}})}f(\Bar{\varphi}) d\Bar{\varphi},
    \vspace{-2mm}
\end{align}
where $\delta$ is the inter-element separation, $\Bar{\varphi}$ represents the angle of a multipath component with a probability density function (PDF) $f(\Bar{\varphi})$. The angle $\Bar{\varphi}$ is expressed as $\Bar{\varphi} = \varphi + \omega$, where $\varphi$ is the central angle, and $\omega$ is a random deviation with zero mean and variance $\sigma_{\varphi}^2$. We assume $\omega$ follows a Laplacian distribution. The parameter, $\sigma_{\varphi}$ is the angular standard deviation (ASD) and decreases as frequency increases. 
\subsubsection{Frequency selective channels}\label{ssec:freq_corr_model}

Since we are dealing with multiple continuous channels over a large bandwidth, it is crucial to account for channel correlations across the frequency range. For this, we incorporate Jake's correlation model for fast-fading Rayleigh channels \cite{Jakes1994}. For simplicity of the explanation, let us consider a single antenna transmitter and a single antenna receiver such that the system becomes Single-Input-Single-Output (SISO). For a SW channel with $L$ equally spaced sub-channels with a bandwidth of $\Delta f$, the $ij^{th}$ element of the fast-fading frequency correlation matrix, $\mathbf{R}_f$ can be expressed as \cite{kongara2004}, 
\begin{align}
    [\mathbf{R}_f]_{ij} &= 1/ \left({1+ 2\pi \Delta f |j-i|\tau_{rms}}\right),
    \label{eqn:jakes_corr}
    \vspace{-3mm}
\end{align}
where $\tau_{rms}$ is the delay spread. 

When generating $L$ frequency-correlated sub-channels, we assume that the spatial and frequency correlations are separable. The resultant channel matrix is $N_{r}\times L$, where columns represent spatial channels at different frequencies. Therefore, for large $L$, calculating and storing the frequency correlation matrix, $\mathbf{R}_f \in \mathbb{C}^{L\times L}$ becomes computationally expensive and memory-intensive. Hence, we propose a blockwise computation method as explained below. 

For large frequency separations, frequency correlations for the corresponding sub-channels are negligible. Therefore, we focus on constructing the correct correlation structure in large blocks of length $2n$. Let $\bm{h}$ be a frequency correlated, unit power channel vector for a single antenna,  of length $2n$  $(<\!\!L)$, with correlation matrix, $\mathbf{R}_f \in \mathbb{C}^{2n\times 2n}$. Then, we can write,
\begin{align}
    \!\!\!\bm{h} &= [\bm{h}_{1}^T, \bm{h}_2^T]^{T}\!\! = \mathbf{R}_f^{\frac{1}{2}}[u_{1},u_{2},\ldots,u_{2n}]^T \!\!\!= \mathbf{R}_f^{\frac{1}{2}}[\bm{u}_{1}^T, \bm{u}_{2}^T]^T,
    \label{eqn:freq_correlated_single_channel}
\end{align}
where $u_{i} \!\sim \! \mathcal{C}\mathcal{N}(0,1)$. We assume that $\bm{h}_1$ and $\bm{h}_2$ have the same correlation pattern as $\mathbf{R}_f$ remains constant across frequency, provided that $\Delta f$ and $\tau_{rms}$ are fixed. Note that $\mathbf{R}_f$ can be written as,
\begin{align}
    \mathbf{R}_f = \begin{bmatrix}
        \mathbf{R}_1 & \mathbf{C} \\
        \mathbf{C}^H & \mathbf{R}_1
    \end{bmatrix}.
    \label{mat:freq_corr_1}
    \vspace{-2mm}
\end{align}
Let the Cholesky decomposition of $\mathbf{R}_f$ be $\mathbf{R}_f = \mathbf{U}^{H}\mathbf{U}$, where $\mathbf{U}$ is upper triangular. Then $\mathbf{R}_f^{\frac{1}{2}} = \mathbf{U}^H$ can be used to generate $\bm{h}$ where $\mathbf{U}^H$ is lower triangular. Now,
\begin{align}
    \mathbf{R}_f &= \begin{bmatrix}
        \mathbf{U}_{1}^H & \mathbf{0} \\
        \mathbf{U}_{2}^H & \mathbf{U}_{3}^H
    \end{bmatrix} \begin{bmatrix}
        \mathbf{U}_{1} & \mathbf{U}_2 \\
        \mathbf{0} & \mathbf{U}_{3}
    \end{bmatrix},
    \label{mat:freq_corr_2}
    \vspace{-2mm}
\end{align}
where $\mathbf{U}_{1}^H$, $\mathbf{U}_{3}^H$ are lower triangular. From (\ref{mat:freq_corr_1}) and (\ref{mat:freq_corr_2}), we can generate $\bm{h}$ in (\ref{eqn:freq_correlated_single_channel}) as,
\begin{align}
    \bm{h} &= [\bm{h}_{1}^T, \bm{h}_2^T]^{T} = \mathbf{R}_f^{\frac{1}{2}}[\bm{u}_{1}^T, \bm{u}_{2}^T]^T \nonumber\\
    &=\begin{bmatrix}
        \mathbf{U}_{1}^{H}\bm{u}_1  \\
        \mathbf{U}_{2}^{H}\bm{u}_1 + \mathbf{U}_{3}^{H}\bm{u}_2
    \end{bmatrix}.
    \label{eqn:h_generation}
\end{align}
We can generalize (\ref{eqn:h_generation}) for any $k$ block of length $n$ as,
\begin{align}
\!\!\!\!\bm{h}_k \!\!=\!\!
\begin{cases}
    \mathbf{U}_{1}^{H}\bm{u}_k, & \text{for } k = 1, \\
    \mathbf{U}_{2}^{H}\left(\mathbf{U}_{1}^{H}\right)^{-1}\bm{h}_{k-1} + \mathbf{U}_{3}^{H}\bm{u}_k, & \text{for } \! k \geq 2,
\end{cases}
\label{eqn:recursive_algo_freq_corr}
\vspace{-2mm}
\end{align}
for all $k \in \left\{1, \ldots, \left\lceil \frac{L}{n}\right\rceil\right\},$ where $\bm{u}_k \!\! \sim \!\mathcal{C}\mathcal{N}(0, \mathbf{I}_n)$. We use (\ref{eqn:recursive_algo_freq_corr}) to generate frequency-correlated channels for each antenna, and then use spatial correlations to create correlations across the antennas. This is achievable if we select an appropriate $n$ such that any two sub-channels separated by at least $2n\Delta f$ are virtually uncorrelated. The formulation in  \eqref{eqn:recursive_algo_freq_corr} has three key advantages: a) it avoids the need to compute and store massive correlation matrices; b) the recursion is simple and fast as the muliplying matrices $\mathbf{U}_{1}$ and $\mathbf{U}_{2}^{H}\left(\mathbf{U}_{1}^{H}\right)^{-1}$ can be pre-computed; and c) it maintains the desired correlation pattern across $2n$ frequencies where $n$ can be chosen to give the desired accuracy.

\subsubsection{Modelling the Rician K-factor}\label{ssec:k_fac}

The Rician K-factor is typically frequency dependent. As the frequency increases, the LoS component becomes more dominant resulting in a higher K-factor. Assuming the K-factor follows a log-normal distribution, as discussed in \cite{mmMagic2017, kfacKristem2018}, we model its mean and variance as functions of frequency, where both increase linearly in the logarithmic domain. Thus, we model the frequency dependent Rician K-factor by a log-normal variable with mean $\mu_{K-dB}$ and variance $\sigma^2_{K-dB}$ where subscript $dB$ indicates that the values are in dB. We present an extrapolation method in Section \ref{Sec:Simulations} to compute $\mu_{K-dB}$ and $\sigma^2_{K-dB}$ as a function of frequency.

\section{Numerical Results}\label{Sec:Simulations}

In this section, we present the SW performance over the new Rician channel model, the impact of MC on the steering vectors in coupled arrays compared to conventional MIMO arrays, and the impact of MC on spatial correlations.
 % and discuss the simulation results for the Super Wideband MIMO channel model (SW MIMO), developed for Rician fading. 
\subsection{Simulation setup}
We consider co-linearly configured ULAs equipped with canonical minimum scattering (CMS) antennas under the Chu constraint, as they provide broader bandwidths \cite{chu1948,kahn1965cms, super_wideband}. The arrays operate between $100$ MHz and $30$ GHz. The source voltage generator's resistance, $R$, and the LNA's internal resistance, $R_{in}$, are both set to $1 \Omega$. Also, the LNA gain, $\beta$, and the noise factor, $N_f$, are set to $10$ and $5$ dB, respectively. The parameter, $\phi$, in (\ref{eqn:cct_theoretic_channel}) which depends on the RLC equivalent circuit model of an antenna element is given as \cite{chu1948},
\begin{align}
    \phi &= \pi - \arctan({2\pi f a_T / c}) - \arctan({2\pi f a_R / c}),
\end{align}
where $a_T$ and $a_R$ represent radii of spheres that enclose a transmit antenna and a receiver antenna, respectively, and $c$ denotes the speed of light. We compute the mutual impedances between the antenna elements using the closed-form expression in \cite{super_wideband} for CMS antennas. Inter-element separation $\delta$ is fixed to $0.5$ cm, and we consider a fixed array size in all scenarios, where MC is controlled by the element sizes $a_T$ and $a_R$. In the figures in this section, tight coupling corresponds to high MC, where the elements are closely packed or connected, and weak coupling corresponds to low MC where the elements are isolated as in conventional MIMO arrays. We consider broadside communication where both the angles of arrival and departure are set to $0$. As in \cite{super_wideband}, we also set the total power to $2$ W, the constant path loss exponent $\gamma$ to $3.5$, and the distance between transmit and receive arrays, $d$, to $90$ m.

We compute $\mu_{K-dB}$ and $ \sigma^2_{K-dB}$ for our developed frequency dependent Rician K-factor in Section \ref{ssec:k_fac} by extrapolating K-factor values from urban micro-cell street canyon measurement campaigns in \cite{mmMagic2017, kfacKristem2018, 3gpp.38.901, kfac2, kfac3, kfac4, kfac5, kfacSamimi2016}. The resultant model of $\mu_{K-dB}$ and $\sigma^2_{K-dB}$ as a function of frequency can be given as,
\vspace{-2mm}
\begin{equation}\label{Eq:lognormMuAndSigma}
\begin{split}
    \mu_{K-dB}&=4.142 \log(f_{GHz}) + 0.246,\\
    \sigma^2_{K-dB}&=0.455 \log(f_{GHz}) + 2.863,
\end{split}
\end{equation}
where $f_{GHz}$ denotes the frequency in GHz.

We first generate a random K-factor at the lowest frequency based on (\ref{Eq:lognormMuAndSigma}) and then standardize the resulting K-factor. We keep that standardized value constant across all frequencies. This is equivalent to a fixed environment so that high or low K-factors at the lowest frequency remain high or low relative to the K-factor distribution across the whole frequency range.

To calculate the spatial correlations, as explained in Section \ref{ssec:spatial_corr}, we adopt a variable ASD which is linearly dropping from $10^\circ$ at the lowest frequency to $5^\circ$ at the highest frequency, to ensure ASD has a decreasing pattern with frequency. Moreover, frequency-correlated scattered channels are generated using the method explained in Section \ref{ssec:freq_corr_model}, with a delay spread $\tau_{rms}$ of $2$ ns.

\subsection{Results for received signal to noise ratio}\label{ssec:snr_results}

Under a transmit power constraint and optimum beamforming, the received signal-to-noise ratio (SNR) of the $i^{th}$ eigenchannel of the SW MIMO system in (\ref{eqn:whitened_mimo_sys}) can be written as,
\begin{equation}
    \text{SNR}_{i}(f) = P_{t,i}(f)\lambda_{i}(\widetilde{\mathbf{H}}^{H}(f)\widetilde{\mathbf{H}}(f)),
\end{equation}
where $P_{t,i}(f)$ is the transmit power allocated for the $i^{th}$ eigenmode of the channel under any power allocation strategy, and $\lambda_{i}(.)$ is the eigenvalue of the $i^{th}$ eigenmode of the channel. 

\begin{figure}[t]
    \centering
    \includegraphics[width=0.51\textwidth]{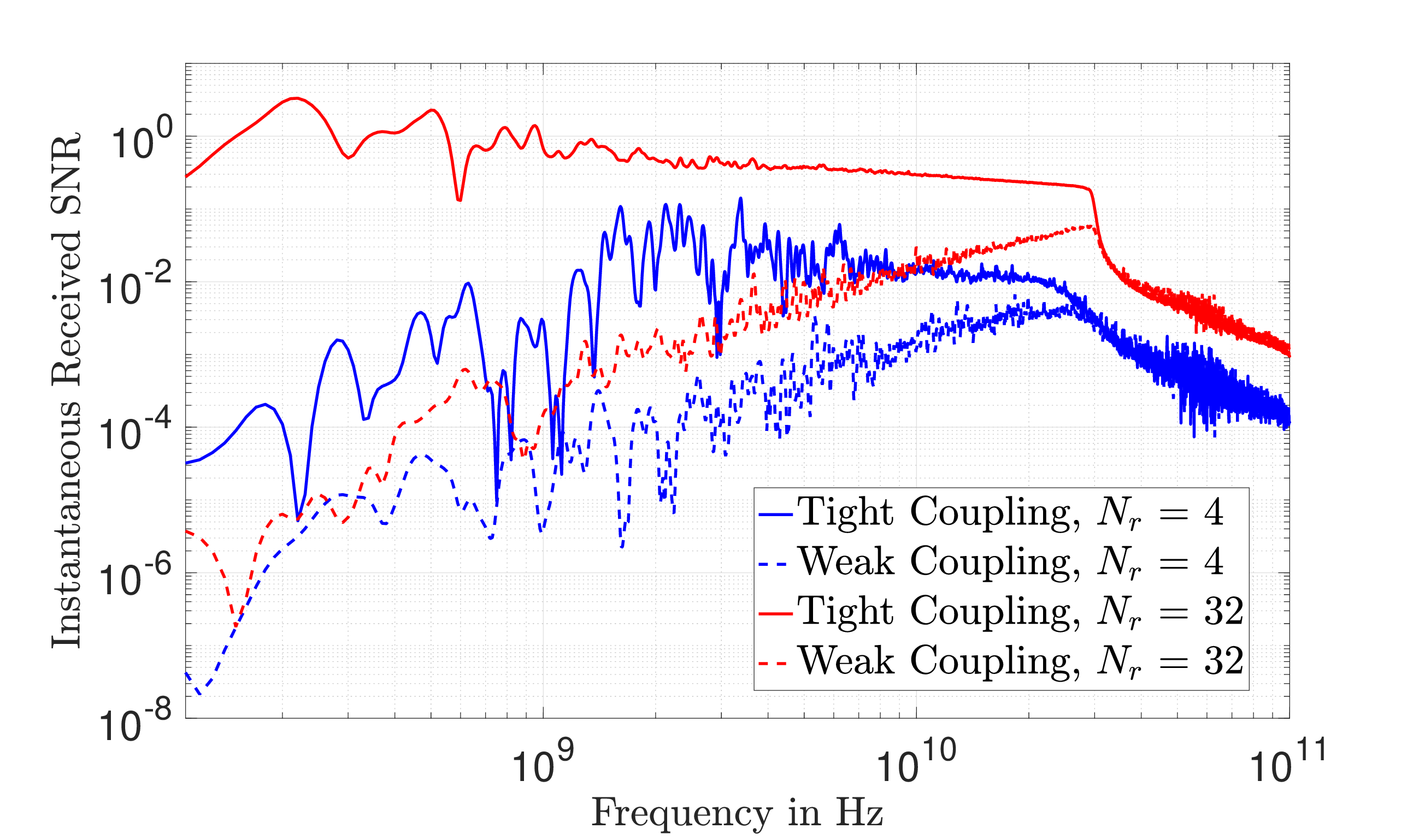}
    \caption{SIMO Instantaneous SNR under equal power allocation for $N_r = {4, 32}$. The transmitter antenna size, $a_T$ is set to $100\delta$ to avoid bandwidth constraints.}
    \label{fig:simo_snr_epa}
\end{figure}

In Fig.~\ref{fig:simo_snr_epa}, we present the instantaneous received SNRs for a SW SIMO system under equal transmit power allocation. We consider received SNRs for $N_r = 4$, and $N_r = 32$ under tight and weak coupling. The figure shows how the SNR profile varies with the frequency. As shown, tight coupling expands the operational bandwidth by leveraging the high MC in the array. This is evident from the high SNR values observed on the left side of Fig.\ref{fig:simo_snr_epa}. This is expected as the array's effective length increases due to high MC, allowing it to behave like a single large antenna at lower frequencies, while individual or smaller groups of elements resonate at higher frequencies. Additionally, having more elements increases the SNR, as more energy is captured. 

The fluctuation of SNR curves in Fig.~\ref{fig:simo_snr_epa} is due to the randomness introduced by the non-line-of-sight components of our channel model. The fluctuation diminishes as frequency increases since the LoS component, which is deterministic, becomes more dominant. 

\subsection{Coupling effects on received scattered power}\label{ssec:rx_power}

\begin{figure}[t]
    \centering
    \includegraphics[width=0.51\textwidth]{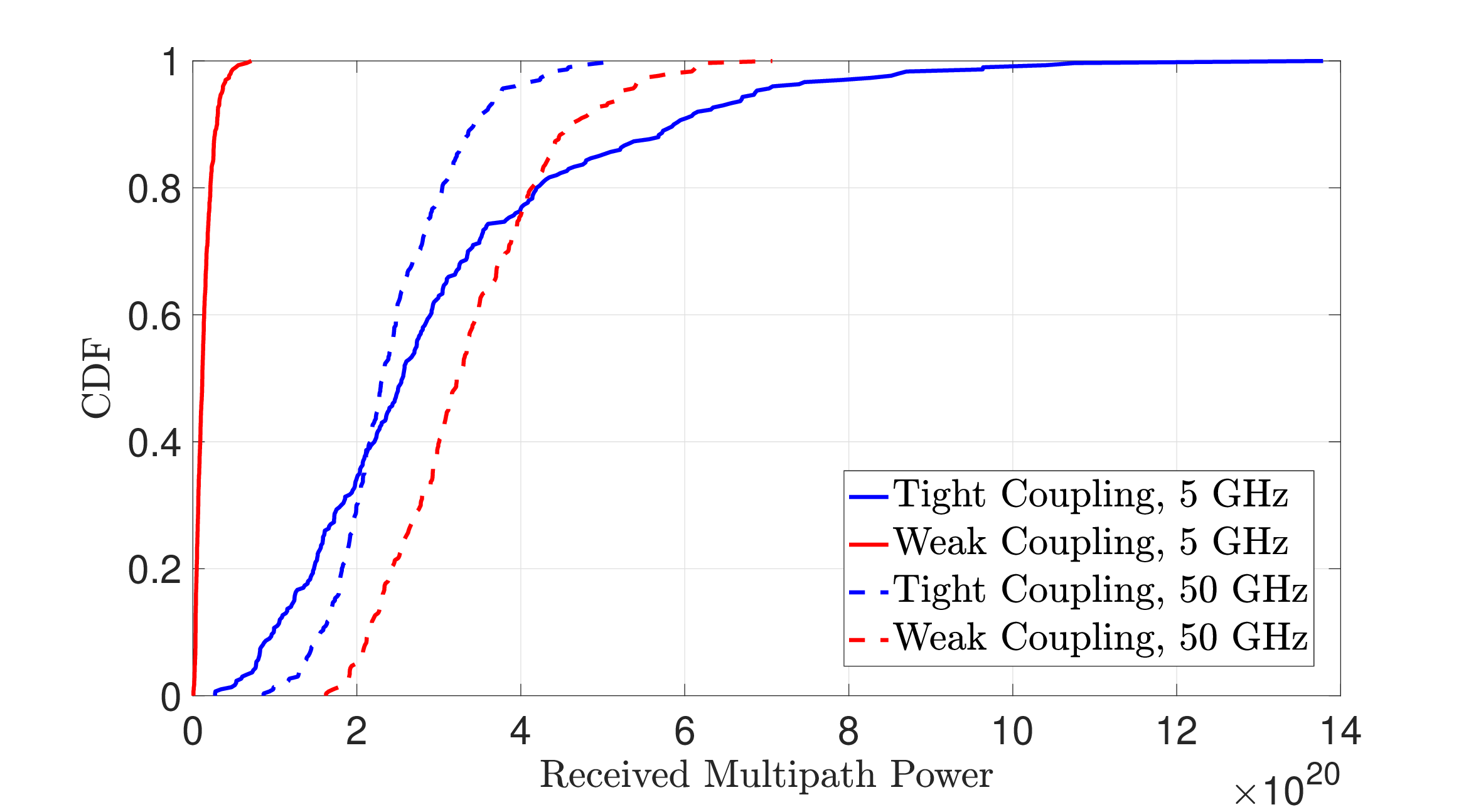}
    \caption{SIMO received signal power CDFs of scattered components at $5$ GHz, and $50$ GHz under tight and weak coupling.}
    \label{fig:power_cdf}
\end{figure}

In Fig.~\ref{fig:power_cdf}, we present the cumulative distribution function (CDF) plots of received scattered powers at $5$ GHz and $50$ GHz. As can be seen in the figure, for $5$ GHz, the high MC that causes the operational bandwidth gain also results in significant received power gains. However, for $50$ GHz low MC outperforms high MC in terms of received power as observed in Fig.~\ref{fig:power_cdf}. This observation is consistent with power loss reported in previous studies that mainly focused on the impact of MC on the antenna elements operating in the resonant frequency such as the works in \cite{mutual_coupling_2018,wallace2004}. However, these studies have not thoroughly investigated the impact of high MC on performance at low frequencies.

\subsection{Coupling effects on the LoS channel}\label{ssec:los_steering_vectors}

\begin{figure}[t] % 'h' for 'here'
    \centering
    \begin{minipage}{0.45\textwidth} % Width of the first image
        \centering
        \includegraphics[width=\textwidth]{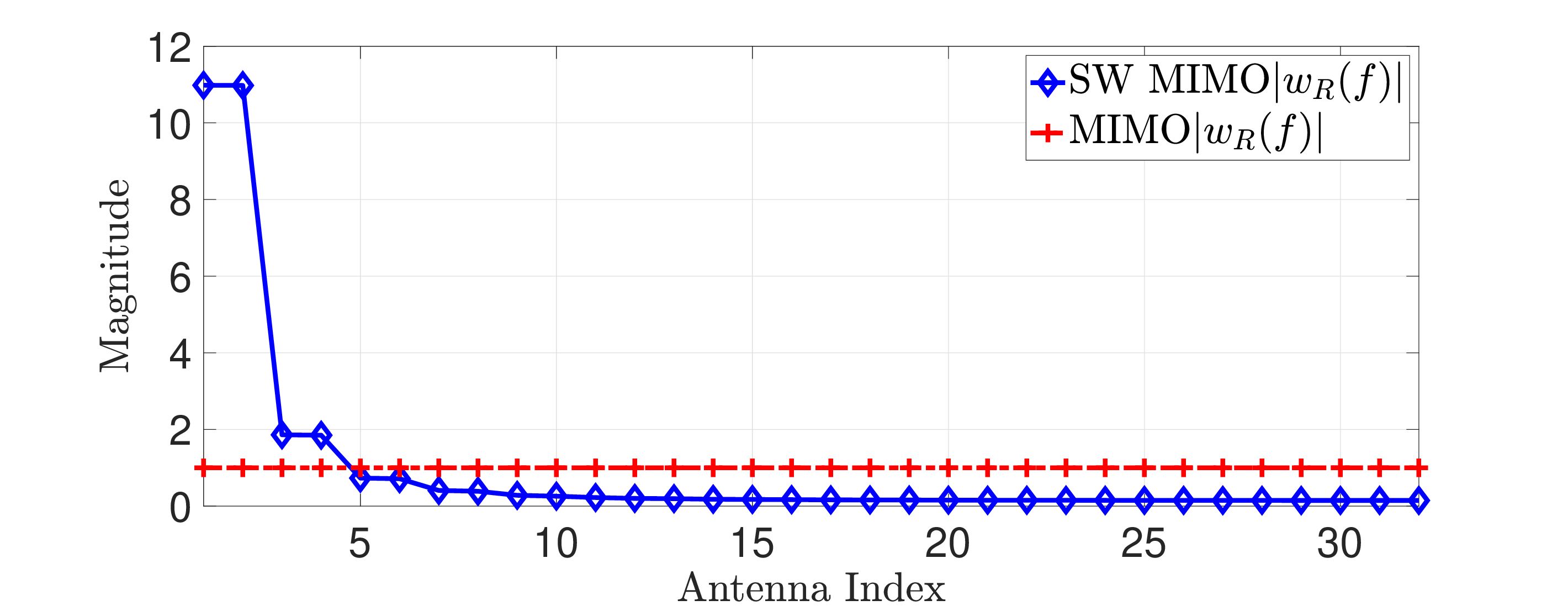}
        \caption*{(a) Receiver Array Steering Vector}
        \label{fig:los_rx_sv}
    \end{minipage}

    \vspace{0.3cm} % Vertical space between the images
    
    \begin{minipage}{0.45\textwidth} % Width of the second image
        \centering
        \includegraphics[width=\textwidth]{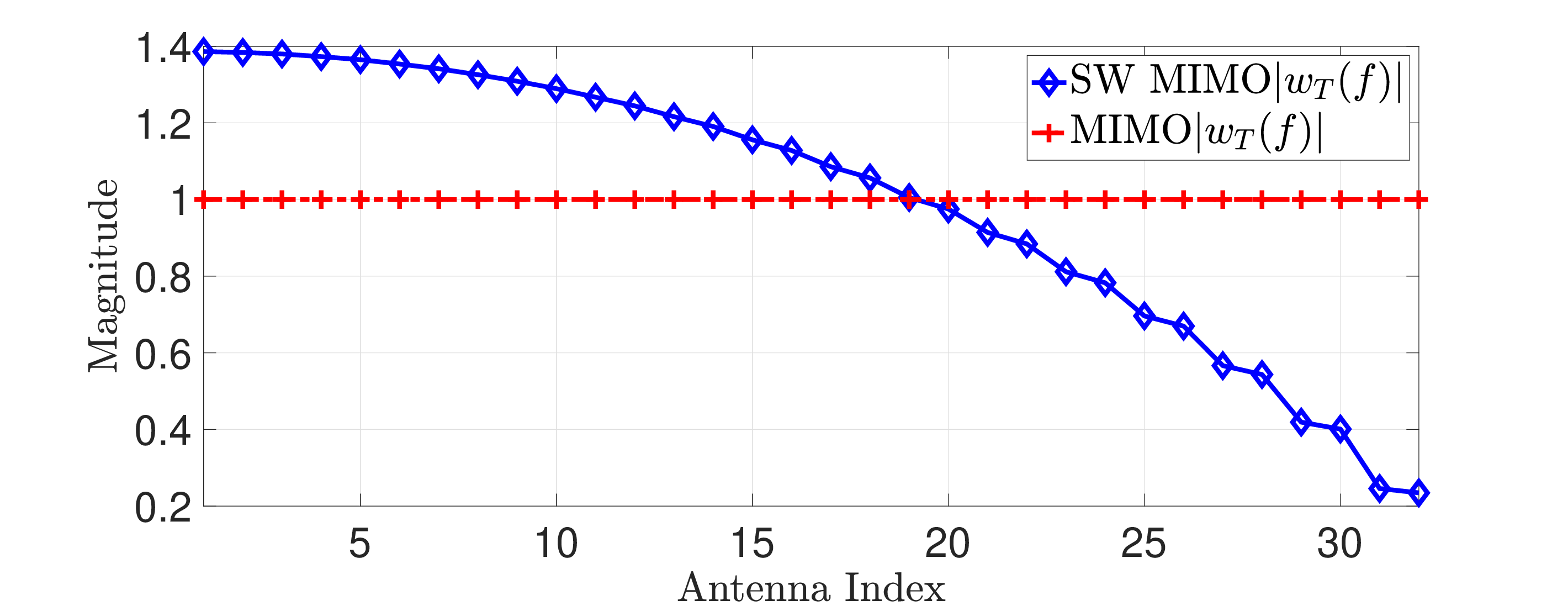}
        \caption*{(b) Transmitter Array Steering Vector}
        \label{fig:los_tx_sv}  
    \end{minipage}
    
    \caption{Ordered, normalized magnitudes of the elements of (a) Receiver, and (b) Transmitter array steering vectors of conventional MIMO and SW MIMO communications for $N_r = N_t = 32$ at $100$ MHz.}
    \label{fig:los_sv}
    \vspace{-3mm}
\end{figure}

In Fig.~\ref{fig:los_sv}, we show the impact of mutual coupling on the steering vectors comprising a LoS channel, for both transmitter and receiver arrays in subfigures \ref{fig:los_sv}(a) and \ref{fig:los_sv}(b), respectively. To do so we compare the ordered, normalized magnitudes of steering vectors for a conventional MIMO array, $\bm{a}_R(.)$ and $\bm{a}_T(.)$, with a SW MIMO array, $\bm{w}_R(.)$ and $\bm{w}_T(.)$, each with 32 antenna elements. Also, all off-diagonal elements of the coupling matrices for the conventional MIMO array are set to zero to eliminate interactions between antenna elements, as conventional MIMO arrays are considered without coupling. As shown in Fig.~\ref{fig:los_sv}, high MC increases the magnitudes of a few elements of the steering vectors, while the majority of the elements decrease so that the magnitudes of all other elements are very small at the considered frequency. This behavior is completely different from the fixed values corresponding to the conventional MIMO array. This amplitude variation means that directional beamforming cannot be achieved with only phase shifts at the antennas and digital processing is required.

\subsection{Coupling effects on spatial correlations} \label{ssec:effects_corr}

\begin{figure}[t]
    \centering \includegraphics[width=0.48\textwidth]{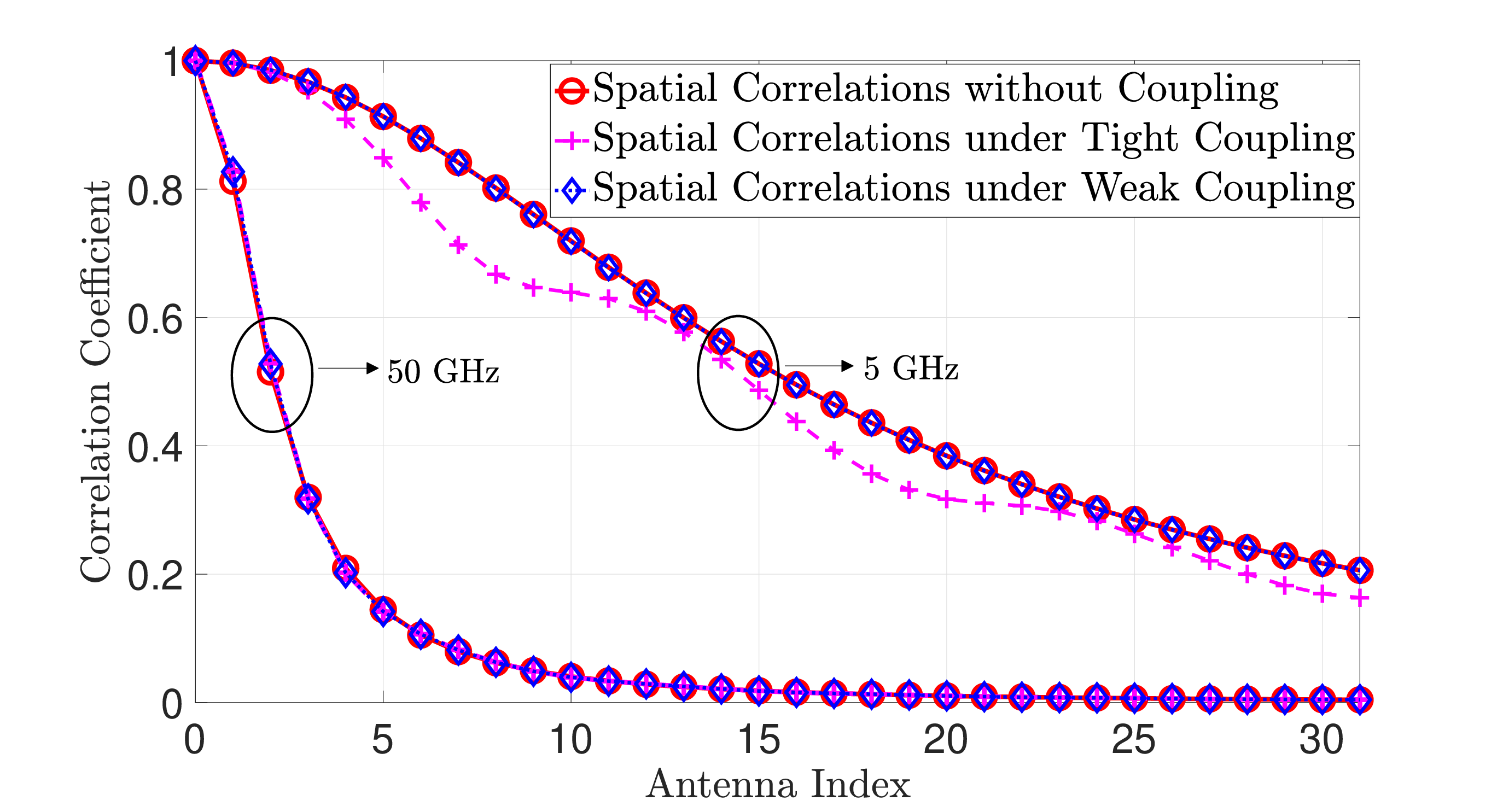}
    \caption{Effective spatial correlations at the receiver with $N_r = 32$, under tight and weak coupling, for $5$ GHz and $50$ GHz.}
    \label{fig:channel_corr}
    \vspace{-6mm}
\end{figure}

The noise correlation and coupling both affect the effective spatial correlations. This is shown in Fig. \ref{fig:channel_corr} where we use (\ref{eqn:equivalent_spatial_correlations}) to plot the values in the first row of the spatial correlation matrix i.e, values corresponding to the correlations of the first antenna element with other elements. The figure shows spatial correlations without coupling generated using the local scattering model discussed in Section \ref{ssec:spatial_corr}, and the effective spatial correlations in (\ref{eqn:equivalent_spatial_correlations}) under tight and weak coupling. 

Based on Fig. \ref{fig:channel_corr} we make the important observation that tight coupling reduces spatial correlations at low frequencies, where bandwidth widening occurs due to high MC. This is due to the combined effect of high coupling and noise correlation in the equivalent correlation matrix $\mathbf{C}_{R}(f)$ in (\ref{eqn:equivalent_spatial_correlations}). Close inspection of $\mathbf{C}_{R}(f)$ shows that the correlations in $\mathbf{R}_{R}(f)$ are partially canceled by the inverses of $\mathbf{R}_{n}(f)$ and the inverse in $\mathbf{P}(f)$. As frequency increases, the influence of high MC on spatial correlations gradually diminishes and the effective spatial correlations become almost identical under both tight and weak coupling. Also, low MC does not significantly impact the effective spatial correlations at any frequency.

\section{Conclusion}\label{Sec:conclusion}

We developed a physically-consistent Rician channel model by accurately accounting for mutual coupling effects by integrating circuit theory into communication theory. We included spatial and frequency correlations in our channel modelling and computed the Rician K-factor across the frequency domain using recent measurement campaigns. We showed that our physically-consistent channel model results in bandwidth-widening, and the simple LoS structure under our channel model behaves differently compared to the conventional antenna arrays. We also highlighted that tight coupling reduces spatial correlations at low frequencies and its impact diminishes for high frequencies. An interesting future extension to this work is analyzing the impact of MC on the performance of multi-user communications systems, under physically-consistent correlated Rician channels.

 \vspace{-1mm}
\bibliographystyle{IEEEtran}
\bibliography{to_arxiv_main}

\begin{thebibliography}{10}
\providecommand{\url}[1]{#1}
\csname url@samestyle\endcsname
\providecommand{\newblock}{\relax}
\providecommand{\bibinfo}[2]{#2}
\providecommand{\BIBentrySTDinterwordspacing}{\spaceskip=0pt\relax}
\providecommand{\BIBentryALTinterwordstretchfactor}{4}
\providecommand{\BIBentryALTinterwordspacing}{\spaceskip=\fontdimen2\font plus
\BIBentryALTinterwordstretchfactor\fontdimen3\font minus \fontdimen4\font\relax}
\providecommand{\BIBforeignlanguage}[2]{{%
\expandafter\ifx\csname l@#1\endcsname\relax
\typeout{** WARNING: IEEEtran.bst: No hyphenation pattern has been}%
\typeout{** loaded for the language `#1'. Using the pattern for}%
\typeout{** the default language instead.}%
\else
\language=\csname l@#1\endcsname
\fi
#2}}
\providecommand{\BIBdecl}{\relax}
\BIBdecl

\bibitem{chataut2020massive}
R.~Chataut \emph{et~al.}, ``{Massive MIMO Systems for 5G and beyond Networks —- Overview, Recent Trends, Challenges, and Future Research Direction},'' \emph{Sensors}, vol.~20, no.~10, p. 2753, 2020.

\bibitem{super_wideband}
M.~Akrout \emph{et~al.}, ``{Super-Wideband Massive MIMO},'' \emph{IEEE J. Sel. Areas Commun.}, vol.~41, no.~8, pp. 2414--2430, 2023.

\bibitem{mutual_coupling_2018}
X.~Chen \emph{et~al.}, ``{A Review of Mutual Coupling in MIMO Systems},'' \emph{IEEE Access}, vol.~6, pp. 24\,706--24\,719, 2018.

\bibitem{wallace2004}
J.~Wallace \emph{et~al.}, ``{Mutual Coupling in MIMO Wireless Systems: A Rigorous Network Theory Analysis},'' \emph{IEEE Trans. Wireless Commun.}, vol.~3, no.~4, pp. 1317--1325, 2004.

\bibitem{munk2006connected}
B.~A. {Munk}, ``{A Wide Band Low Profile Array of End Loaded Dipoles with Dielectric Slab Compensation},'' in \emph{EuCAP 2006}, ser. ESA Special Publication, H.~{Lacoste} \emph{et~al.}, Eds., vol. 626, Oct. 2006, p.~9.

\bibitem{connected_arrays}
D.~Cavallo, ``\BIBforeignlanguage{English}{{Connected Array Antennas : Analysis and Design}},'' Phd Thesis 1 (Research TU/e / Graduation TU/e), Electrical Engineering, 2011.

\bibitem{Ivrlac_main}
J.~A. Nossek \emph{et~al.}, ``{Towards A Circuit Theory of Communication},'' in \emph{2009 IEEE Int. Conf. Electromagn. Adv. Appl.}, 2009, pp. 750--753.

\bibitem{matlab_mc_model}
Y.~Wu \emph{et~al.}, ``{Effects of Antenna Mutual Coupling on the Performance of MIMO Systems},'' in \emph{Proc. 29th Symp. Inf. Theory Benelux}, 2008.

\bibitem{shyianov2022}
V.~Shyianov \emph{et~al.}, ``{Achievable Rate With Antenna Size Constraint: Shannon Meets Chu and Bode},'' \emph{IEEE Trans. Commun.}, vol.~70, no.~3, pp. 2010--2024, 2022.

\bibitem{warnick2009}
K.~F. Warnick \emph{et~al.}, ``{Minimizing the Noise Penalty Due to Mutual Coupling for a Receiving Array},'' \emph{IEEE Trans. Antennas Propag.}, vol.~57, no.~6, pp. 1634--1644, 2009.

\bibitem{emil2017}
E.~Bj\"{o}rnson \emph{et~al.}, ``{Massive {MIMO} Networks: {Spectral}, Energy, and Hardware Efficiency},'' \emph{Found. Trends Signal Process.}, vol.~11, no. 3-4, pp. 154--655, 2017.

\bibitem{Jakes1994}
W.~C. Jakes, Ed., \emph{{Microwave Mobile Communications}}.\hskip 1em plus 0.5em minus 0.4em\relax New York, NY, USA: IEEE Press, 1994.

\bibitem{kongara2004}
K.~P. Kongara \emph{et~al.}, ``{Performance Analysis of Adaptive MIMO OFDM Beamforming Systems},'' in \emph{2008 IEEE Int. Conf. Commun.}, 2008, pp. 4359--4365.

\bibitem{mmMagic2017}
K.~Haneda \emph{et~al.}, ``{Measurement Results and Final mmMAGIC Channel Models},'' Tech. Rep., 05 2017.

\bibitem{kfacKristem2018}
V.~Kristem \emph{et~al.}, ``{Outdoor Wideband Channel Measurements and Modeling in the 3–18 GHz Band},'' \emph{IEEE Trans. Wireless Commun.}, vol.~17, no.~7, pp. 4620--4633, 2018.

\bibitem{chu1948}
L.~J. Chu, ``{Physical Limitations of Omni-directional Antennas},'' \emph{J. Appl. Phys.}, vol.~19, no.~12, pp. 1163--1175, 1948.

\bibitem{kahn1965cms}
W.~Kahn \emph{et~al.}, ``{Minimum-Scattering Antennas},'' \emph{IEEE Trans. Antennas Propag.}, vol.~13, no.~5, pp. 671--675, 1965.

\bibitem{3gpp.38.901}
{3GPP}, ``{Study on channel model for frequencies from 0.5 to 100 GHz (Release 14)},'' Tech. Rep. {3GPP TR 38.901 Release 14}, June 2017.

\bibitem{kfac2}
D.~Shakya \emph{et~al.}, ``{Radio Propagation Measurements and Statistical Channel Models for Outdoor Urban Microcells in Open Squares and Streets at 142, 73, and 28 GHz},'' \emph{IEEE Trans. Antennas Propag.}, vol.~72, no.~4, pp. 3580--3595, 2024.

\bibitem{kfac3}
S.~Zhu \emph{et~al.}, ``{Probability Distribution of Rician $K$-Factor in Urban, Suburban and Rural Areas Using Real-World Captured Data},'' \emph{IEEE Trans. Antennas Propag.}, vol.~62, no.~7, pp. 3835--3839, 2014.

\bibitem{kfac4}
S.~Sun \emph{et~al.}, ``{Millimeter Wave Small-scale Spatial Statistics in an Urban Microcell Scenario},'' in \emph{2017 IEEE Int. Conf. Commun. (ICC)}, 2017, pp. 1--7.

\bibitem{kfac5}
M.~D. Balde \emph{et~al.}, ``{A 32 GHz Urban Micro Cell Measurement Campaign for 5G Candidate Spectrum Region},'' in \emph{2017 11th Eur. Conf. Antennas Propag. (EuCAP)}, 2017, pp. 1803--1807.

\bibitem{kfacSamimi2016}
M.~K. Samimi \emph{et~al.}, ``{Local Multipath Model Parameters for Generating 5G Millimeter Wave 3GPP-like Channel Impulse Response},'' in \emph{2016 10th Eur. Conf. Antennas Propag. (EuCAP)}, 2016, pp. 1--5.

\end{thebibliography}

\end{document}